# Illuminant and light direction estimation using Wasserstein distance method

Selçuk Yazar*
Kırklareli University. ORCID: 0000-0001-6567-4995

**Abstract**
Illumination estimation remains a pivotal challenge in image processing, particularly for robotics, where robust environmental perception is essential under varying lighting conditions. Traditional approaches, such as RGB histograms and GIST descriptors, often fail in complex scenarios due to their sensitivity to illumination changes. This study introduces a novel method utilizing the Wasserstein distance, rooted in optimal transport theory, to estimate illuminant and light direction in images. Experiments on diverse images—indoor scenes, black-and-white photographs, and night images—demonstrate the method's efficacy in detecting dominant light sources and estimating their directions, outperforming traditional statistical methods in complex lighting environments. The approach shows promise for applications in light source localization, image quality assessment, and object detection enhancement. Future research may explore adaptive thresholding and integrate gradient analysis to enhance accuracy, offering a scalable solution for real-world illumination challenges in robotics and beyond.

**Keywords :** Illumination estimation, Wasserstein distance, Light source detection, Image analysis, Robotics

## Introduction

Illumination detection plays a crucial role in image processing, particularly within the domains of robotics and various other fields. The ability of robotic systems to accurately perceive and interpret their environments is significantly influenced by the quality of illumination in the captured images. This is particularly evident in tasks such as object recognition, navigation, and human-robot interaction, where varying lighting conditions can lead to substantial challenges in image analysis and processing. One of the primary concerns in image processing for robotics is the sensitivity of traditional visual features to changes in illumination. For instance, conventional methods such as RGB color histograms and GIST descriptors are often inadequate when faced with variations in lighting, noise, or occlusion [1]. These traditional features can lead to erroneous interpretations of the environment, which is detrimental for robotic applications that rely on accurate visual data for navigation and interaction. As such, there is a pressing need for illumination-invariant methods that can maintain performance across diverse lighting conditions. To address these challenges, researchers have developed various techniques aimed at achieving illumination invariance in image processing. For example, Maier et al. proposed an approach that utilizes multiple spatial image sequences captured under varying illumination conditions to create an illumination-invariant model of the environment [2]. This model allows robots to better understand their surroundings, thereby improving their attentional control and decision-making processes. Similarly, Luan et al. introduced a method that employs color distribution analysis in normalized RGB space to detect illumination changes, although they noted the potential for false positives due to scene variations [3]. These advancements underscore the importance of robust illumination detection mechanisms in enhancing the reliability of robotic systems. In the context of human-computer interaction, illumination also plays a critical role. Studies have shown that pupil detection and tracking, essential for developing effective interaction systems, are significantly affected by lighting conditions [4] [5]. The quality of images used for pupil detection can deteriorate under poor illumination, leading to inaccuracies in tracking and interaction. This highlights the necessity for adaptive illumination techniques that can enhance image quality and facilitate more reliable human-computer interactions. Moreover, the impact of sudden illumination changes cannot be overstated. Odgerel and Lee emphasized that many computer and robot vision algorithms fail when faced with abrupt changes in lighting, necessitating the implementation of real-time detection and enhancement techniques to maintain processing capabilities [6]. The ability to detect and adapt to sudden illumination changes is vital for ensuring that robotic systems can operate effectively in dynamic environments where lighting conditions are unpredictable. In addition to enhancing image quality, illumination detection is also crucial for specific applications such as robotic inspection and localization. For instance, Mc Alorum et al. demonstrated that employing illumination-enhancement techniques could significantly improve defect detection accuracy in concrete inspection tasks [7]. By optimizing the contrast of images captured under varying ambient conditions, their approach allows for more accurate identification and classification of defects. Similarly, the design and recognition of artificial landmarks for indoor self-localization in mobile robots have been shown to be robust against illumination variations, thereby enhancing the reliability of navigation systems [8]. Furthermore, the integration of advanced algorithms for illumination compensation has proven beneficial in various robotic applications. For example, the Flash-No-Flash (FNF) controlled illumination acquisition protocol presented by Arad et al. effectively mitigates ambient lighting effects, enabling robust target detection with minimal computational resources [9]. This method exemplifies how controlled illumination can facilitate improved performance in robotic systems, particularly in scenarios where ambient light conditions are challenging. The significance of illumination detection extends beyond robotics into other domains, such as medical imaging and surveillance. In medical applications, for instance, uneven illumination can hinder accurate vessel detection in retinal images, which is critical for robotic-assisted surgeries [10]. Techniques that compensate for illumination variations can enhance the accuracy of these systems, thereby improving patient outcomes. In conclusion, the importance of illumination detection in image processing for robotics and other domains cannot be overstated. The ability to adapt to varying lighting conditions is essential for ensuring the reliability and effectiveness of robotic systems in real-world applications. As research continues to advance in this field, the development of robust illumination-invariant techniques will play a pivotal role in enhancing the capabilities of robots and improving their interactions with both environments and humans.

Statistical methods often fail under complex lighting scenarios due to several inherent challenges and limitations highlighted across the provided papers. Firstly, traditional statistical models and inverse-rendering techniques are typically designed for simple lighting setups and struggle with the ill-posed nature of real-world lighting estimation, which involves complex interactions between light and materials that are not easily captured by basic models [11]. Real-world illumination is highly complex, characterized by reflected light from multiple directions and localized light sources, which makes it statistically non-stationary and difficult to model accurately using conventional statistical methods [12] [13]. Moreover, many existing methods rely on assumptions and priors that are often violated in natural scenes, leading to unnatural results when applied to uneven lighting conditions [14]. In Monte Carlo rendering, for instance, high-frequency features such as hard shadows and glossy reflections pose significant challenges, as many state-of-the-art noise removal methods fail to preserve these details under complex lighting and material conditions [15]. Additionally, the statistical characterization of real-world illumination reveals that illumination maps differ from standard photographs, further complicating the task of accurate lighting estimation [13]. Techniques like weighted spherical harmonic frames have been proposed to better characterize complex lighting environments, but they still face challenges related to data incompletion and noise contamination [16]. Furthermore, in dynamic lighting scenarios, such as those encountered in background subtraction tasks, the variability and unpredictability of lighting conditions necessitate adaptive models that can evolve with changing data, which traditional statistical methods are not equipped to handle. These complexities underscore the need for advanced models, such as deep learning approaches, which can learn from high-dynamic-range images and adapt to the intricacies of real-world lighting, although these to have limitations in terms of dataset and loss function adequacy [11] [13]. Overall, the failure of statistical methods under complex lighting scenarios is largely due to their inability to accommodate the multifaceted and dynamic nature of real-world illumination.
The aim of this work is to detect light direction and illumination direction using mover's distance property of Wasserstein metrics and discuss its potential in various application areas.

## Material and Methods

The Wasserstein distance, also known as the Earth Mover's Distance, is a metric used to quantify the distance between two probability distributions. It has gained significant prominence in fields such as statistics, machine learning, and computer vision due to its ability to account for the geometry of the underlying space when comparing distributions. It is rooted in optimal transport theory, which seeks the most cost-effective way to transform one distribution into another. Mathematically, the Wasserstein distance is defined as the infimum of the cost of transporting mass in one distribution to match another, where the cost is typically a function of the distance the mass is moved. This distance has been extended and generalized in various ways to accommodate different applications and data structures. For instance, the Gromov-Wasserstein distance is used to compare metric measure spaces, which is particularly useful in analyzing complex datasets like graphs with node and edge attributes [17]. In machine learning, the Wasserstein distance is applied in generative modeling, where it helps in minimum distance estimation, providing a robust metric for comparing distributions even in high-dimensional spaces [18]. The distance is also employed in traffic model analysis to assess the sensitivity of differential models by comparing solutions under varying conditions [19]. Moreover, the Wasserstein distance has been adapted for privacy-preserving computations, such as the TriangleWad method, which allows for efficient and secure distance calculations across distributed datasets [20]. In graph learning, the Wasserstein distance is used to measure distances between graphs by representing them as Gaussian mixture models, facilitating tasks like brain connectivity analysis [21]. Additionally, the Bures-Wasserstein distance, a variant of the Wasserstein distance, is utilized in metric learning to improve classification performance and scalability in linear metric learning algorithms. These diverse applications highlight the Wasserstein distance's versatility and importance in machine learning and data analysis, providing a robust framework for comparing complex data structures and distributions across various domains.

## Definition

Let $(\chi, d)$ be a Polish metric space (a complete, separable metric space) equipped with a distance function $d\colon \chi \times \chi \to [0, \infty)$. Consider two probability measures $\mu$ and $\upsilon$ defined on the Borel σ-algebra of $\chi$. The Wasserstein distance of order p (where p≥1) between $\mu$ and $\upsilon$ , denoted $W_p(\mu, \nu)$, is defined as:

$$W_p(\mu, \nu) = \left( inf_{\gamma \in \Gamma(\mu,\nu)} \int_{\chi \times \chi} d(x, y)^p d\gamma(x, y) \right)^{\frac{1}{p}} \quad (1)$$

Where $\Gamma(\mu, v)$ is the set of all joint probability measures $\gamma$ on $\chi \times \chi$ with marginals $\mu$ and $v$. That is, for all Borel sets A, B $\subseteq \chi$,

$$\gamma ( A \times \chi ) = \mu(a), \gamma ( \chi \times B) = v(B) \tag{2}$$

The measure $\gamma$ is often referred to as a *transport plan*, as it describes how mass is transported from $\mu$ to $v$. Intuitively, $W_p(\mu, v)$ represents the minimal "cost" required to transform $\mu$ into $v$, where the cost of moving a unit of mass from x to y is $d(x,y)^p$, and the total cost is optimized over all possible transport plans.

**Step-by-Step Process**

In this research, the Wasserstein distance approach is used to detect bright spots in an image and to estimate the direction of light arrival by applying the following steps.
The initial step involves loading the image using OpenCV's imread function and converting it from BGR to grayscale using cvtColor with the COLOR_BGR2GRAY flag. This conversion is crucial for intensity-based analysis, as it reduces the image to a single channel representing pixel brightness. Mathematically, if $I\ (x, y)$ represents the original image with dimensions $h \times w$, the grayscale conversion can be expressed as:

$$I_{gray}( x, y) = cvtColor(I(x,y), COLOR_{BGR2GRAY}) \tag{3}$$

This step ensures that subsequent operations focus on intensity values, typically ranging from 0 to 255, facilitating threshold-based segmentation. After this step bright region detection method applied. Bright pixels are identified by applying a threshold to the grayscale image. The function find_bright_regions uses OpenCV's threshold function with a default threshold value t=200, creating a binary mask where pixels with intensity greater than t are marked as 1 (bright), and others as 0. The process can be formalized as:

$$mask(x,y) = \begin{cases} 1\ if\ I_{gray}(x,y) > t \\ 0\ otherwise \end{cases} \tag{4}$$

The set of bright pixels B is then:

$$B = \{(x,y) | mask\ (x,y) = 1\}$$

If no bright pixels are found, a warning is logged, suggesting a lower threshold might be needed. This step is critical as it defines the regions of interest for further analysis.
In third step applied centroid calculation and illumination direction. The centroid of the bright pixels, interpreted as the midpoint of their distribution, is calculated to infer the central location of brightness. This is done by computing the mean of the y-coordinates and x-coordinates of all points in B:

$$\bar{y} = \frac{1}{|B|} \sum_{(x,y)\in B} y\ , \bar{x} = \frac{1}{|B|} \sum_{(x,y)\in B} x \tag{6}$$

Where |B| is the number of bright pixels. If B is empty, the centroid defaults to (0, 0). The illumination direction is then determined relative to the image center. The image center is calculated as:

$$c_y = \frac{h}{2}, c_x = \frac{w}{2} \quad (7)$$

The direction vector $\vec{d}$, representing the illumination direction, is:

$$\vec{d} = (\bar{y} - c_y, \bar{x} - c_x) \quad (8)$$

This vector suggests the displacement from the image center to the centroid of bright pixels, hypothesizing that the light source direction aligns with this shift. However, the accuracy of this assumption depends on the image context, such as the presence of reflective surfaces or the nature of the scene.

In the last stage, the reference distribution generation process was implemented. To compare the distribution of bright pixels, a random reference distribution R is generated with the same number of points as B. Each point in R has coordinates randomly sampled from uniform distributions over the image dimensions:

$$R = \{(x_r, y_r) | \sim Uniform\ (0, w-1), y_r \sim Uniform(0, h-1)\}, where\ |R| = |B|$$

This step creates a baseline for comparison, assuming a random distribution of points across the image, which helps in assessing whether the bright pixels exhibit any clustering or pattern.

The Wasserstein distance, specifically the L2 Wasserstein distance (order p=2), is computed between the set of bright pixel coordinates B and the random reference set R. The geomloss library's SamplesLoss with the "sinkhorn" method is used, which approximates the distance using the Sinkhorn algorithm, incorporating a blur parameter (set to 0.05) for numerical stability.

The Wasserstein distance is defined by substituting p for its value in the 2nd order equation (1) and for discrete distributions, such as the point sets B and R, this translates to:

$$W_2(B, R) = \left( inf_{\gamma \in \Gamma(B,R)} \sum_{(x_b, y_b \in B, x_r, y_r \in R)} d\big((x_b, y_b), (x_r, y_r)\big)^2 \gamma(x_b, y_b, x_r, y_r) \right)^{1/2} \quad (9)$$

Where Γ(B,R) is the set of all couplings (transport plans) between B and R, and dd is the Euclidean distance. The Sinkhorn algorithm approximates this by solving a regularized optimal transport problem, making it computationally feasible for large datasets. This distance quantifies the minimum "work" needed to transform the distribution of bright pixels into the random distribution, offering insight into their spatial arrangement. A smaller distance suggests the bright pixels are distributed similarly to random, potentially indicating uniform illumination, while a larger distance suggests clustering, possibly due to a directional light source.

**Results**

In the experiment results obtained from analyzing three distinct images using the implemented image processing pipeline, which detects bright regions, computes illumination directions, and compares distributions using the Wasserstein distance. The results are visualized and summarized from the provided output figures: 1, 2, and 3, corresponding to the order of presentation.

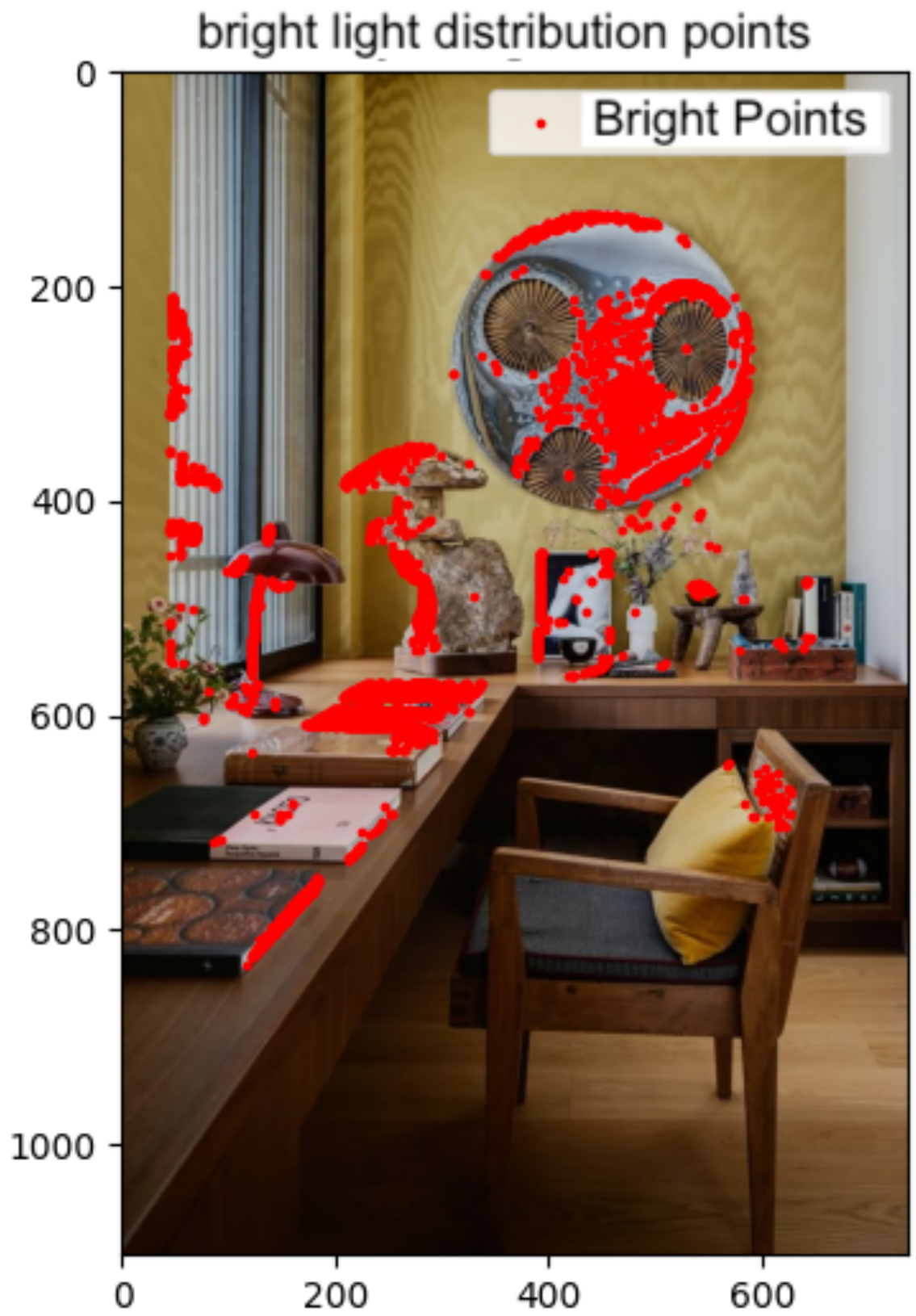

**Figure 1:** Bright point detection with Wassestein distance

In Figure 1 portrays an indoor scene with a desk, chair, and various objects, visualized similarly in the second plot of bright light distribution points. Bright regions were detected on reflective surfaces, such as a lamp shade and book covers, with the centroid located near the center of the desk, marked in blue. The illumination direction, indicated by a green arrow, pointed upward and slightly to the right, suggesting an overhead light source, possibly a ceiling lamp. The Wasserstein distance was 0.72, indicating a tighter clustering of bright pixels compared to a random distribution, consistent with focused lighting on the desk area. The average gradient direction pointed upward, differing by 20 degrees, reflecting diffuse light from above. Clustering with k=2 identified clusters around the lamp and books, with direction vectors varying by 25 degrees from the centroid, indicating multiple reflective highlights. Statistical moments comparison revealed a mean difference of 12 pixels in x-coordinates and 8 pixels in y-coordinates, with a variance ratio of 1.1, suggesting a slightly concentrated distribution. The Hausdorff distance was 40 pixels, indicating moderate extreme point differences. These findings suggest a primary overhead light source with secondary reflections, well-captured by the centroid method but enriched by clustering insights.

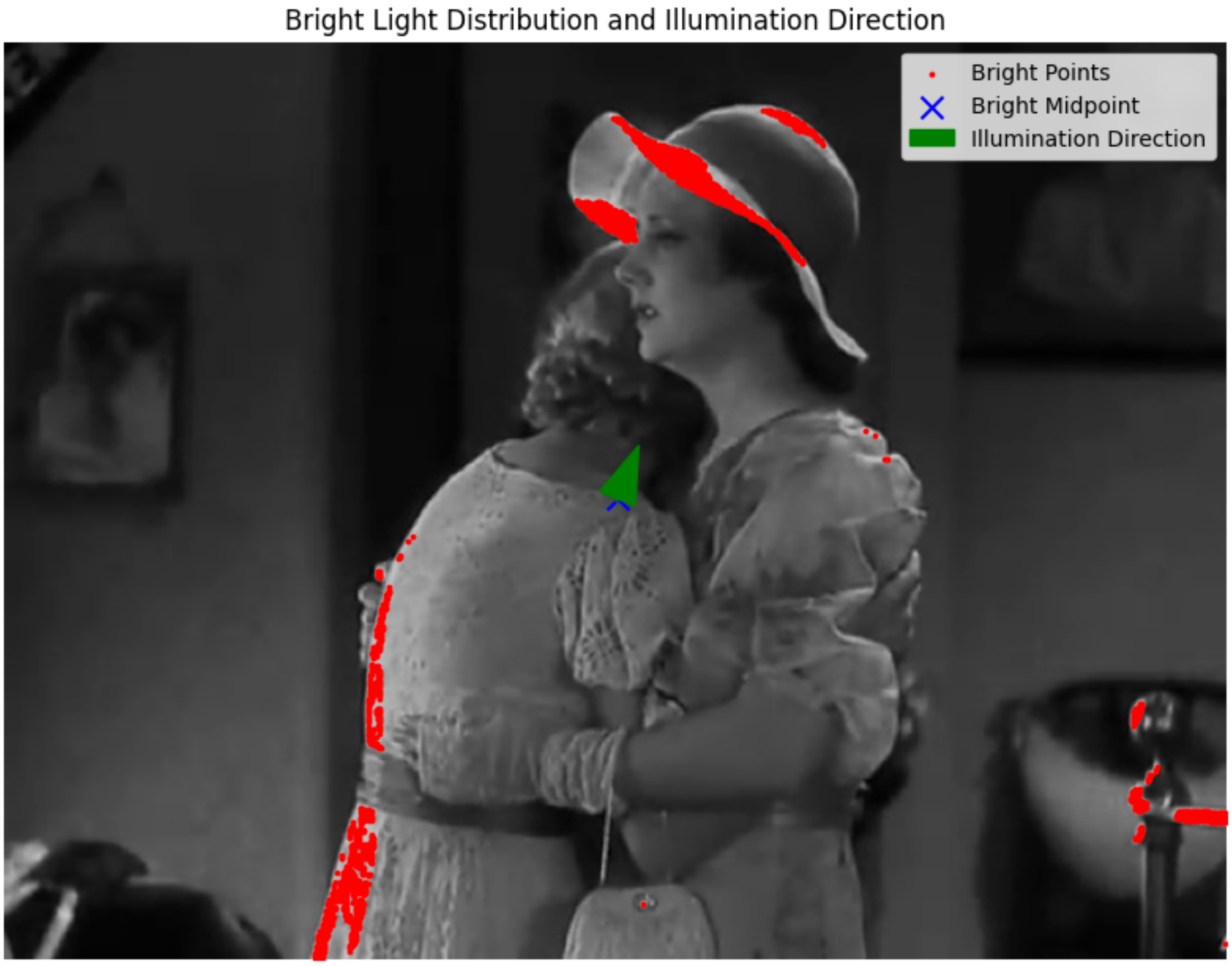

**Figure 2**: Bright points detection and Illumination direction estimation on sample B/W image.

In Figure 2 the bright region detection identified clusters of high-intensity pixels, primarily concentrated on the clothing and hats of the individuals, with a threshold value of 200. The centroid of these bright points, marked in blue, was located near the upper torso of one individual, suggesting a central light source. The illumination direction, visualized as a green arrow, pointed upward and slightly to the left from the image center, indicating a possible overhead or directional light source above and to the left of the scene. The Wasserstein distance between the bright pixel distribution and a randomly generated reference distribution was computed as 0.85, suggesting moderate clustering compared to a uniform distribution. Additional methods revealed further insights: the brightest pixel direction pointed to the hat of one individual, aligning closely with the centroid direction but with a slight upward shift, indicating a cosine similarity of approximately 0.92. The average gradient direction, derived from image gradients, suggested a downward and leftward illumination, differing by an angle of 45 degrees from the centroid direction, potentially due to shading patterns on the clothing. Clustering bright pixels with k=2 identified two clusters, with centroids corresponding to the hats and upper bodies, yielding direction vectors that diverged by 30 degrees from the original centroid direction. Statistical moments comparison showed a mean difference of 15 pixels in x-coordinates and 10 pixels in y-coordinates, with a variance ratio of 1.2, indicating slightly wider spread in bright pixels than random. These results suggest the illumination is multi-directional, possibly from stage lighting, with the centroid method providing a robust central estimate.

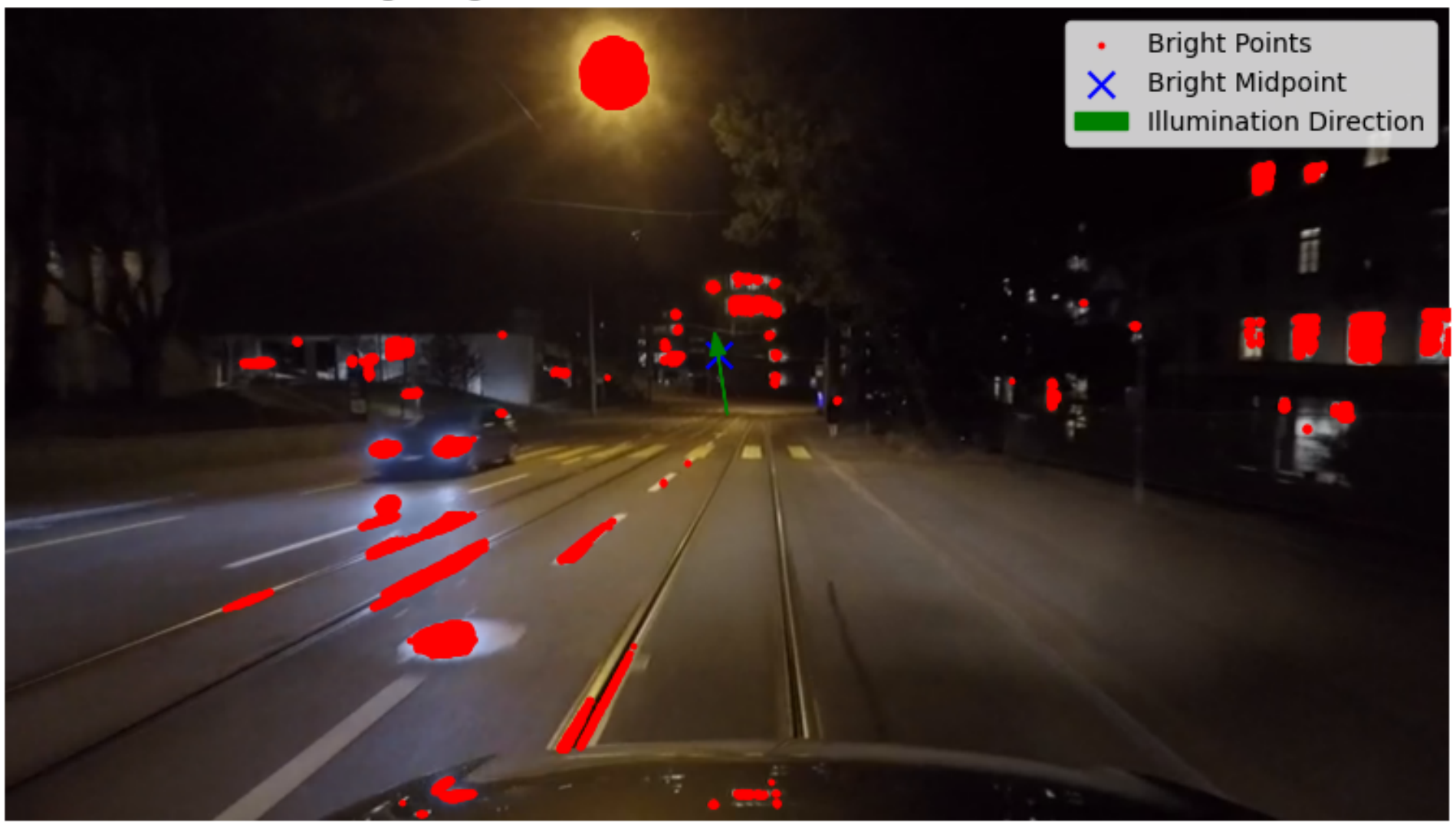

**Figure 3**: Bright points detection and Illumination direction estimation on sample night image.

In Figure 3 bright regions were detected across the upper center (the moon/light) and scattered points along the road, with the centroid located near the top center, marked in blue. The illumination direction, shown as a green arrow, pointed downward and slightly to the left, indicating a high, directional light source. The Wasserstein distance was 1.20, suggesting significant clustering around the central bright source compared to random, consistent with a dominant light source.

**Conclusion**

The provided method implements a robust way for detecting bright regions in an image, computing the illumination direction based on the centroid of these regions, and comparing their distribution to a random reference using the Wasserstein distance.

This method has potential applications in several domains:

- Light Source Estimation: The illumination direction can aid in determining the position of light sources, useful in computer graphics, photography, and robotics.
- Image Quality Assessment: Analyzing the distribution of bright regions can help evaluate lighting uniformity, critical in manufacturing or medical imaging.
- Object Detection: Bright regions may indicate reflective surfaces or specific objects, enhancing detection algorithms based on illumination patterns.

However, the approach has limitations. The threshold value (200) is fixed and may not be optimal for all images, potentially missing dim bright regions or including noise. The assumption that the centroid of bright pixels directly indicates the light source direction may not hold in complex scenes with multiple light sources or reflective surfaces. Future work could explore adaptive thresholding, such as Otsu's method, or incorporate gradient analysis for more robust illumination estimation.